\documentclass[twocolumn]{aastex631}
\usepackage{lipsum, babel}
\usepackage{xspace}

\newcommand{\FindingCRDs}{\texttt{FindingCRDs}\xspace}
\newcommand{\FitKinPA}{\texttt{fit\_kinematic\_pa}\xspace}
\newcommand{\FindPeaks}{\texttt{find\_peaks}\xspace}

\shorttitle{\FindingCRDs on MaNGA DR17}
\shortauthors{Piper et al.}

\begin{document}
\graphicspath{{figures/}}

\title{A New Identification Method and Sample of Counter-Rotating Disk Galaxies in SDSS-IV MaNGA DR17}

\author{Maxwell Piper}
\affiliation{Reed College, 3203 SE Woodstock Boulevard, Portland, OR 97202, USA}
\affiliation{W. M. Keck Observatory, 65-1120 Mamalahoa Hwy, Waimea, HI 96743, USA}
\email{mpiper@keck.hawaii.edu}

\author{Alison Crocker}
\affiliation{Reed College, 3203 SE Woodstock Boulevard, Portland, OR 97202, USA}
\email{crockera@reed.edu}

\begin{abstract}
Counter-Rotating Disk (CRD) galaxies have two co-spatial stellar disks rotating in opposite directions, and provide a rare opportunity to study external gas accretion and dynamical assembly processes. In the 16th data release of the Mapping Nearby Galaxies at Apache Point Observatory (MaNGA) survey, only 64 CRDs were visually identified. Using this as a training sample, we developed an automated pre-selection method that reduces the number of galaxies requiring visual inspection by removing systems unlikely to host counter-rotation. Applying this method to MaNGA Data Release 17, we identified 126 confirmed CRDs and an additional 143 candidate galaxies, more than doubling the MaNGA CRD sample. With this extended sample, we analyzed their Baldwin, Phillips, and Terlevich (BPT) emission-line diagrams and compared them with a matched control sample of early-type galaxies (ETG). We found no statistically significant difference in photoionization sources between CRDs and the ETG control sample, indicating emission-line diagnostics cannot solely be used to identify counter-rotating disks, nor do they correspond to a distinct present-day photoionization signature. Our method facilitates efficient discovery of CRDs in large spectroscopic surveys, enabling stronger statistical studies of their formation and evolution.
\end{abstract}


\section{Introduction} \label{sec:intro}

Counter-Rotating Disk (CRD) galaxies are galaxies that contain two co-spatial, co-planar stellar disks rotating in opposite directions. The first early-type barred CRD galaxy, NGC 4546, was identified in 1987, where its gas and stars rotated with similar amplitudes but in opposite directions \citep{Galletta1987}. Shortly after, NGC~4550 was also detected to have two counter-rotating stellar disks \citep{Rix1992, Rubin1992}, and has been heavily used as a cosmological laboratory for galaxy formation and evolution. CRD galaxies represent a rare class in which a significant fraction of the stellar population has formed through a different mechanism than the rest. In particular, they allow us to study constraints on gas accretion and merger history in a galaxy's formation, ultimately probing how galaxies build and evolve their stellar disks. 

Multiple formation pathways have been proposed for CRDs through detailed analysis of individual galaxies. NGC~4550 likely formed through a major-merger with a particular configuration \citep{Crocker2009, Johnston2013, Puerari_2001}. However, most CRDs likely formed when gas accretion that eventually undergoes star-formation settles into a galaxy’s midplane and
then makes a younger stellar disk. \citep{Bevacqua_2021}. \citet{Lu2021} found this accretion can originate from mergers under specific initial conditions. Through IllustrisTNG simulations, \citet{Nelson2018} found that counter-rotators experience relatively few merger events overall. However, when a merger occurs with orbital angular momentum opposite to that of the progenitor galaxy, it is much more likely to form a CRD. In fact, a merger event with a gas-rich dwarf companion galaxy meeting the required angular momentum constraints is much more likely to form a counter-rotator than that of a gas-poor system \citep{Thakar1997}. 

The overall sample of CRD galaxies is relatively small, with
approximately a hundred having been classified. Their rarity may stem from the intrinsic difficulty in building such counter-rotating structures, the observed inclination of the galaxy, or instrument resolution. Of this small sample, nearly all were observed with large Integral Field Unit (IFU) spectroscopic surveys such as the SDSS-IV Mapping Nearby Galaxies at Apache Point Observatory (MaNGA) \citep{Law2015}. IFU surveys are ideal for finding CRDs due to their large sample sizes and their ability to spatially resolve stellar and gas kinematics. However, the current CRD census remains limited.

\begin{figure} [t]
\centering
\begin{tabular}{c}
\includegraphics[width = 0.82\linewidth]{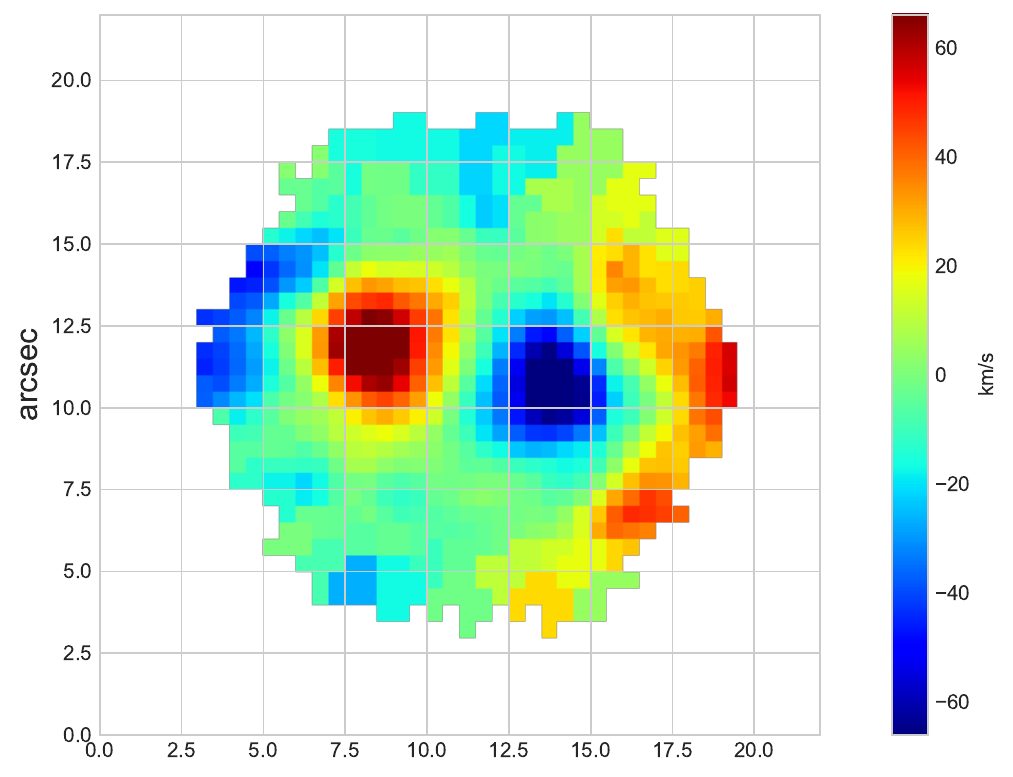} \\
\includegraphics[width = 0.82\linewidth]{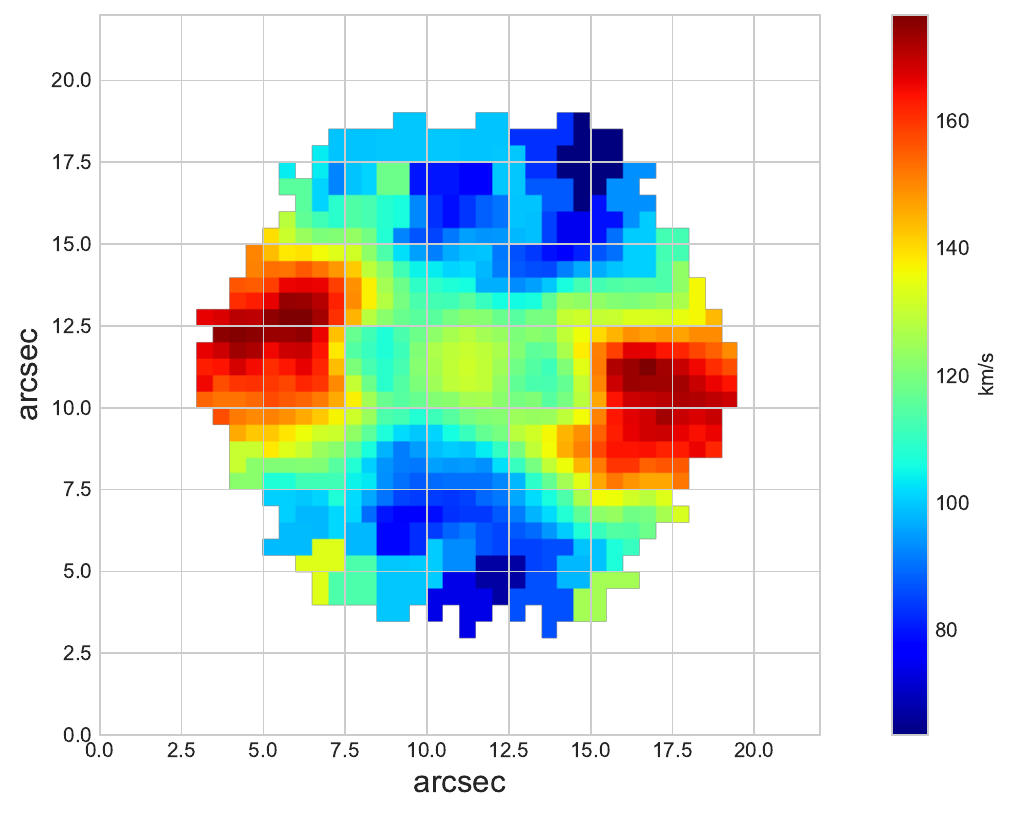}
\end{tabular}
\caption{Stellar velocity ($V_\star$; top) and stellar velocity dispersion ($\sigma_\star$; bottom) maps for MaNGA ID 1-136248 from the MaNGA DAP. The $V_\star$ map exhibits two redshifted and two blueshifted regions characteristic of counter-rotation, while the $\sigma_\star$ map exhibits two spatially distinct peaks associated with the overlapping stellar disks.}
\label{fig: example CRD}
\end{figure}

Using the MaNGA IFU survey, CRDs are identified through visual inspection of their mean stellar velocity ($V_\star$) and stellar velocity dispersion ($\sigma_\star$) maps, examples of which can be seen in Fig.~\ref{fig: example CRD}. In $V_\star$, counter-rotation produces two pairs of redshifted and blueshifted regions on either side of the central bulge. In $\sigma_\star$, counter-rotation produces two spatial regions where both stellar disks are present, resulting in two large peaks (often called ``2$\sigma$" galaxies) that are separate from that produced by the central bulge. Further, if counter-rotation is present in $V_\star$, the spatial pixels (spaxels) that correspond to the large 2$\sigma$ peaks in $\sigma_\star$ often read $\approx0$ km/s in $V_\star$. This denotes the region where the two stellar disks contribute comparable flux. To either side of this area are the specific regions where the inner and outer disks dominate, producing the distinct CRD pattern in $V_\star$. Given these intrinsic characteristics, no scalable method currently exists to identify CRDs in modern IFU surveys.

Recently, \citet{Bevacqua_2021} visually identified a sample of 64 CRD galaxies in MaNGA DR16 \citep{MaNGADR16}. However, given the recent rise in IFU surveys, these surveys are getting larger and larger, making visual inspection of all released galaxies incredibly cumbersome. Furthermore, due to the small sample size, meaningful statistics from CRD population studies aimed at how galactic stellar disks evolve are difficult. This motivated the creation of a new identification tool that helps the CRD identification process in large IFU surveys, allowing a larger sample of CRDs to be created which will enable a better characterization of their formation channels.

In this work, we present a Python program, \FindingCRDs, that partially automates the visual inspection process, reducing the set of galaxies requiring visual inspection to those that show potential counter-rotation. In Section~\ref{sec: sample}, we outline the sample selection criteria and the method of \FindingCRDs. In Section~\ref{sec: results}, we apply \FindingCRDs to DR17 and show its results. In Section~\ref{sec: gas}, we use the new CRD sample from Section~\ref{sec: results} to investigate their photoionization sources and compare these with a matching set of Early Type Galaxies (ETG) to test a new possible identification metric. Finally, in Section~\ref{sec: conclusion} we conclude the paper. 

\section{Data and Methods} \label{sec: sample}

\subsection{Data}

MaNGA DR17 consists of 10,145 galaxy datacubes, 5524 of which  not included in DR16. Each galaxy's stellar kinematics were measured from spatially binned spectra using the MaNGA Data Analysis Pipeline (DAP). The DAP performs Voronoi binning with a minimum signal-to-noise ratio (S/N) of 10 and fits stellar and gas kinematics using the penalized Pixel Fitting Method (pPXF, \citet{cappellari2006}) using the MILES Hierarchical Clustering stellar template library \citet{Westfall_2019}. For each bin, the default DAP configuration utilizes a two-iteration pPXF fit using an additive 8th order Legendre polynomial for the continuum and a Gaussian line-of-sight velocity distribution for the stellar and gas kinematics. After analyzing all Voronoi bins, the DAP produces a multi-extension MAPS file for each galaxy containing the stellar and gas kinematics for each spaxel. These MAPS allow visual inspection of the stellar velocity ($V_\star$), stellar velocity dispersion ($\sigma_\star$), and H$\alpha$ velocity fields. Additionally, the DAP provides an overarching ``DAPall" catalog containing global and fit-specific metrics for all galaxies.

\subsection{The \FindingCRDs\ Algorithm} \label{subsec: the algorithm}

To reduce the need for visual inspection of all MaNGA DR17 galaxies, we developed \FindingCRDs, an automated pre–selection tool designed to identify galaxies that may host counter–rotating stellar disks. Rather than fully replacing visual classification, the goal of \FindingCRDs\ is to eliminate galaxies that clearly lack counter–rotation, thereby reducing the number of systems requiring manual inspection. 


\subsubsection{Training sample and parameter tuning} \label{subsubsec: sample and tuning}

\FindingCRDs parameters were tuned using the visually confirmed DR16 CRD sample of \citet{Bevacqua_2021}. \FindingCRDs has two important caveats with respect to the pure visual-inspection work of \citet{Bevacqua_2021}. First, because the 2D velocity map is reduced to a 1D extraction along the kinematic position angle (PA$_{\text{kin}}$), kinematically decoupled cores are very likely to be included. In a visual inspection, these are more likely to be excluded by rejecting objects with a warp in their velocity field. Second, we do not attempt to include galaxies with extended, high velocity dispersions as was done in \citet{Bevacqua_2021}. These objects would be hard to distinguish with our methods and would require inspection of their photometric data to ensure the extended velocity dispersion peak is not due to a bar.

These caveats require us to exclude four galaxies with extended peaks in $\sigma_\star$. Additionally, we excluded: four galaxies whose PA$_{\text{kin}}$ could not be fit, two nearly face–on galaxies lacking clear peaks in $\sigma_\star$, and three borderline CRDs with ambiguous kinematic signatures. This left a training sample of 51 confirmed CRDs.

For a galaxy to be passed to \FindingCRDs, we imposed three global requirements:

\begin{enumerate}
\item Galaxies must contain at least 60 Voronoi bins.
\item A signal–to–noise floor of S/N = 4 was required for velocity bins.
\item A signal–to–noise floor of S/N = 6 was required for velocity dispersion bins.
\end{enumerate}

These thresholds were empirically calibrated to be the largest values that still successfully recovered the full 51 CRD training sample.

\subsubsection{Algorithm workflow} \label{subsubsec: workflow}

\FindingCRDs begins by determining PA$_{\rm kin}$ using \FitKinPA \citep[Appendix C of][]{Krajnovi2006}. Stellar velocity and velocity dispersion measurements are then extracted from the spaxels within the Voronoi along this axis. Only spaxels meeting the S/N thresholds described above are retained.

\paragraph{Velocity }
\ \FindingCRDs first searches for evidence of counter–rotation in $V_\star$. CRDs are expected to show a velocity flip when moving along PA$_{\rm kin}$ from the outer disk to the inner disk on either side of the galaxy center. Local extrema in the velocity profile are identified using \FindPeaks with parameters empirically calibrated on the training sample (height = 0 km\,s$^{-1}$, width = 2 spaxel pairs, prominence = 2 km\,s$^{-1}$).

To avoid spurious detections, extrema are retained only if they contain at least two unique Voronoi bins with distinct velocity measurements. A galaxy is flagged as showing clear counter–rotation if at least three high–quality velocity extrema are detected across the two halves of the galaxy.

\paragraph{Velocity dispersion }
\ \FindingCRDs then searches for the characteristic double–peaked $\sigma_\star$ profile associated. Peaks are identified using \FindPeaks with parameters scaled to each galaxy (height = $\langle\sigma_\star\rangle + \frac{1}{5}\mathrm{std}(\sigma_\star)$ km\,s$^{-1}$, width = 1.9 spaxels, prominence = 5 km\,s$^{-1}$).

Peaks associated with the central bulge (identified via the flux maximum) are removed. Similar to the velocity criteria, the remaining velocity dispersion extrema must contain at least two independent measurements above the detection threshold to be considered reliable. A galaxy is flagged as having a ``2$\sigma$" velocity dispersion profile if two high–quality peaks are found on opposite sides of the galaxy center.

\paragraph{Final classification }
\ Galaxies satisfying either the velocity or velocity dispersion criterion are flagged as potential CRDs and saved for visual inspection. During visual inspection, systems with especially clear signatures are classified as confirmed CRDs, while borderline cases (e.g., possible kinematically decoupled cores or near face–on systems) are labeled as ``maybe CRDs'' (mCRD) to allow for classification uncertainty.

At its core, \FindingCRDs\ serves as a pre–selection tool that efficiently narrows the MaNGA sample to galaxies most likely to host counter–rotating stellar disks.

\section{Results on DR17} \label{sec: results} \label{subsec: new sample}

\begin{figure*} [t]
\centering
\begin{tabular}{cc}
\includegraphics[width = 0.5\linewidth]{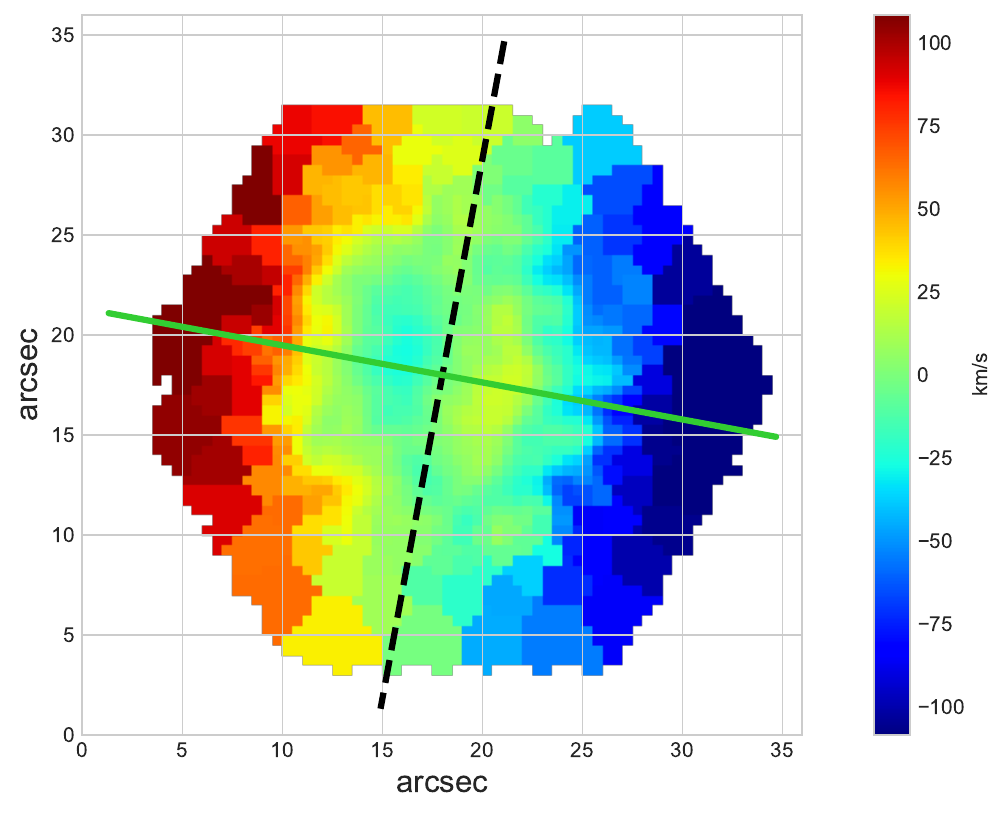} 
\includegraphics[width = 0.5\linewidth]{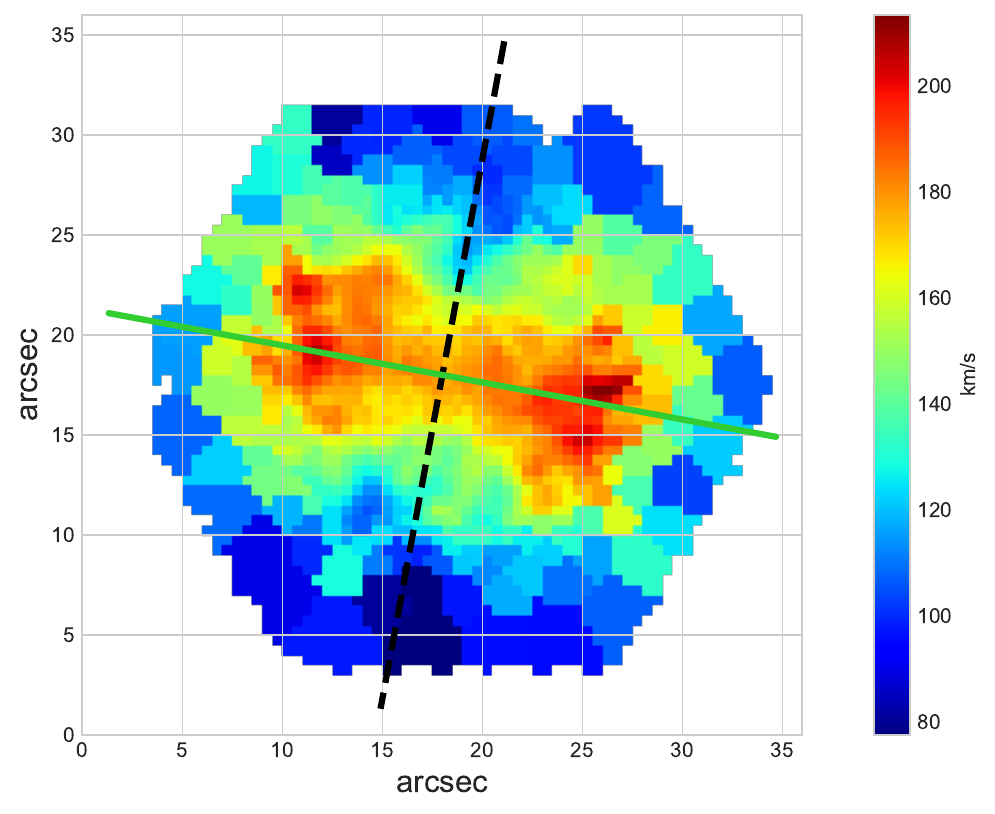}\\
\includegraphics[width = 0.485\linewidth]{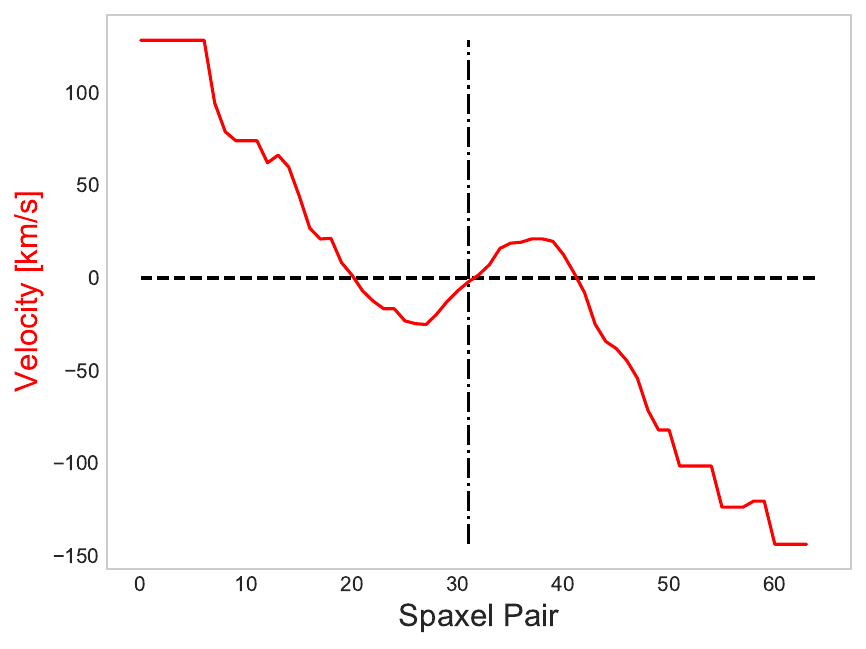} 
\includegraphics[width = 0.515\linewidth]{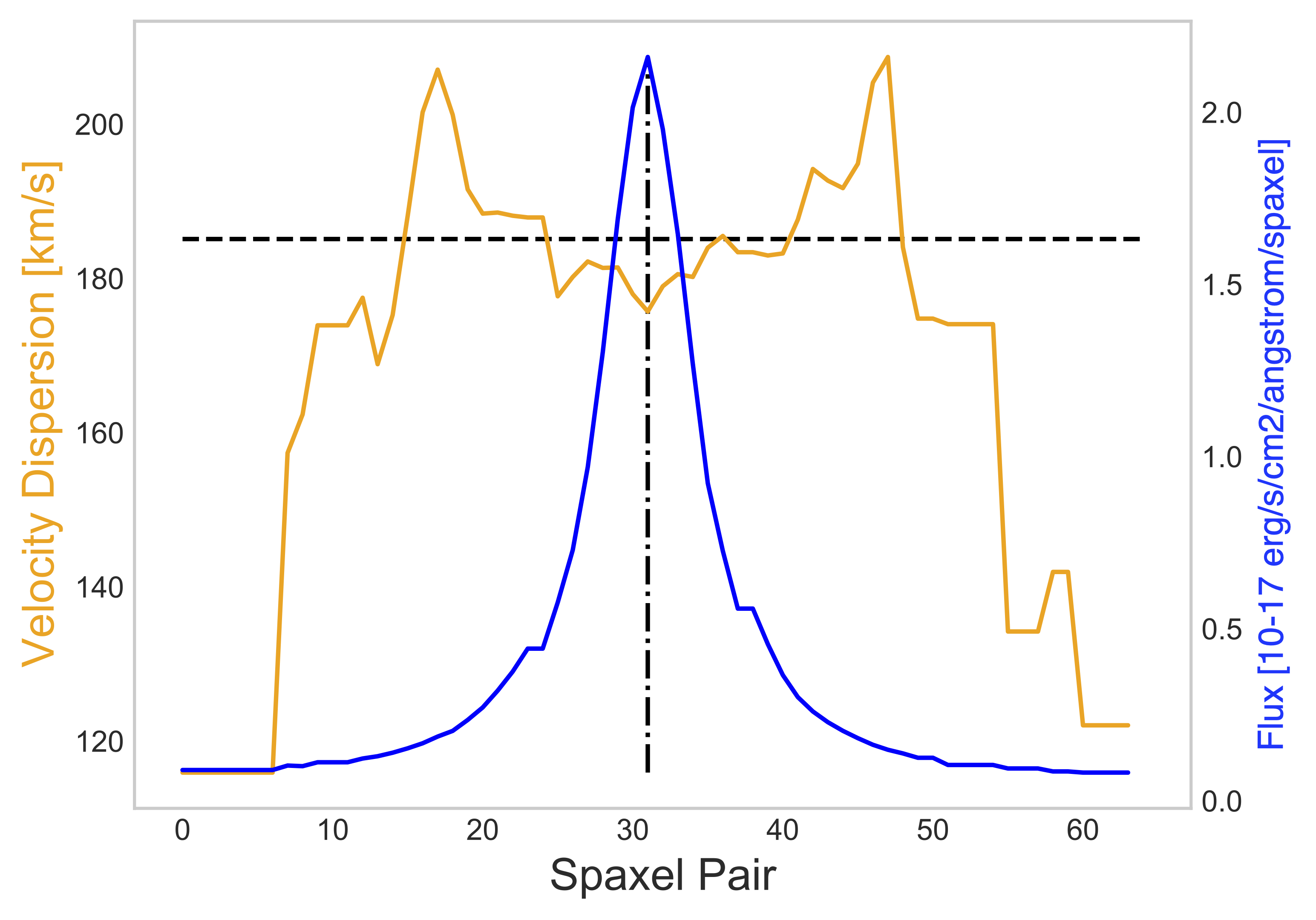}
\end{tabular}
\caption{ MaNGA ID 1-78381. \emph{Left} Plots of the full $V_\star$ with PA$_{\text{kin}}$ superimposed and the recorded velocity of the spaxel pairs along PA$_{\text{kin}}$. \emph{Right} Plots of the full $\sigma_\star$ with PA$_{\text{kin}}$ super imposed and the recorded velocity dispersion of the spaxel pairs along PA$_{\text{kin}}$. The positive velocity extrema were flagged in spaxel pairs 0-13 and 33-40, and the negative velocity extrema were flagged in spaxel pairs 23-30 and 55-58. The sigma extrema were flagged in spaxel pairs 14-23 and 41-48. The dashed horizontal line is the height used in \FindPeaks, and the dot-dashed vertical line is the central spaxel pair. }
\label{fig: finding_crds.py results 10515-12702}
\end{figure*}

Of the 10,145 MaNGA DR17 galaxies, 1,801 were flagged as potential counter-rotators, reducing the number of galaxies requiring visual inspection by $\approx$ 85\%. After visual inspection, 126 galaxies were classified as confirmed CRDs and an additional 143 were classified as mCRDs. This corresponds to a confirmation rate of $\approx$ 7\%. While this success fraction may be modest, our method is intentionally tuned for high completeness rather than high purity. \FindingCRDs prioritizes the recovery of true CRDs while dramatically reducing the total manual inspection workload. 

An example of a flagged galaxy can be seen in Fig.~\ref{fig: finding_crds.py results 10515-12702}, where $V_\star$ and $\sigma_\star$ are shown with PA$_{\text{kin}}$ superimposed in green. Below are plots of the recorded velocity and velocity dispersion of the spaxels along PA$_{\text{kin}}$, where the vertical dot-dashed line is the central spaxel pair and the horizontal dashed lines are the \FindPeaks heights. It is clear both from $V_\star$, $\sigma_\star$, and these plots that this galaxy has both velocity counter-rotation and a clear $2\sigma$ distribution, therefore this is a newly classified MaNGA DR17 CRD galaxy. This example illustrates how \FindingCRDs flags galaxies with the signature kinematic structures of counter-rotation prior to manual visual confirmation.

The 143 mCRDs represent borderline systems whose primary identification feature is summarized in Table ~\ref{tab: mCRD}. These include: kinematically decoupled cores, near face-on galaxies, systems with marginal velocity extrema, and galaxies with weak double-peaked velocity dispersion profiles. Because visual confirmation remains necessary for CRD identification, some level of classification disagreement is unavoidable. We therefore retain these galaxies as a supplementary sample to enable a flexible sample selection for future studies.

For galaxies that were flagged by \FindingCRDs but ultimately rejected,  many exhibited moderately large velocity dispersion along PA$_{\text{kin}}$ that mimicked $2\sigma$ peaks but were not associated with genuine counter-rotation. Increasing the $\sigma_\star$ \FindPeaks threshold would reduce these false positives, but at the cost of missing confirmed CRDs. We argue overshooting the number of galaxies that are flagged for visual inspection is better than excluding a fair number of true CRDs at the benefit of not needing to visually inspect as many.

\subsection{Sample Statistics} \label{subsec: stats}

Basic galaxy parameters are summarized in Table~\ref{tab: dr17 info}. Column 2's ``TTYPE" is used as a proxy for the galaxy's morphology from the MaNGA Deep Learning Morphology (MDLM-VAC) Value Added Catalogs \citep{Fischer2019}. The TTYPE can range form -3 to 10, where TTYPE $<$ 0 corresponds to ellipticals, TTYPE $>$ 0 spirals, and TTYPE $\approx$ 0 lenticulars.  Column 5's ``$\log_{10}\sigma_{\text{e}}$" is the velocity dispersion within a circular aperture spanning out to the galaxy's effective radius on a log scale. Column 6's ``Detection" refers to which CRD feature was flagged in \FindingCRDs. Column 7 and 8 refer to the position angle of the kinematic semi-major axis for the mean stellar velocity and H$\alpha$ map, respectively. Finally, Column 9's ``Gas Alignment" refers to which stellar disk the gaseous disk is corotating with. 

\citet{Fischer2019} also provide a ``Visual Classification" parameter for specifying galaxy morphology. Using this, our final CRD sample consisted of 48 ellipticals, 36 lenticulars, and 42 spirals. The sample of DR17 we considered consisted of 2529 ellipticals, 1598 lenticulars, and 6156 spirals. This suggests CRDs constitute 1.9\% of ellipticals, 2.3\% of lenticulars, and 0.7\% of spirals, which is consistent with the findings of \citet{Bevacqua_2021} (being $<5\%$, $<3\%$, and $<1\%$ respectively). This confirms that CRDs remain more probable with early-type galaxies \citep{Corsini2014}, although their presence in spirals demonstrates that counter-rotation is not restricted to quiescent systems.

\citet{Bevacqua_2021} describe 8 of their 64 CRDs to be ``CRD in formation" due to their blueish colors, disturbed kinematic maps, and irregular shapes in the SDSS image. Using their sample, we find the that largest $g-r$ absolute magnitude (proxy for `blueness') to be 0.72. Applying this criterion to our extended sample, 37 CRDs fall below this threshold and exhibit similarly blueish SDSS images and disrupted kinematics. These galaxies are labeled with $\S$ in column 1 of Table~\ref{tab: dr17 info} and represent promising candidates for recently or actively forming CRDs. 

Overall, the application of \FindingCRDs to MaNGA DR17 more than doubles the known CRD population within the survey. This expanded sample enables statistically meaningful comparisons between CRDs and control galaxy populations, which we explore in the following section.

\section{Ionized Gas} \label{sec: gas}

It is thought the predominant formation scenario for a CRD is the retrograde accretion of gas that eventually undergoes star-formation, suggesting the progenitor disk counter-rotates with respect to the gaseous disk \citep{Corsini2014}. Because this gas is expected to be actively forming stars at the time of accretion, its emission-line spectrum should be dominated by hydrogen recombination lines produced by HII regions. Since this sample of MaNGA CRDs is so new, little analysis has been done to investigate whether star formation ionization sources in CRDs differ systematically from those in non-CRD galaxies. If such differences exist, nebular emission could provide an additional observational signature of counter-rotation. 

\subsection{CRD Ionization Sources} \label{subsec: CRD sources}

\begin{figure*} [t]
\centering
\begin{tabular}{ccc}
\includegraphics[width = 0.318\linewidth]{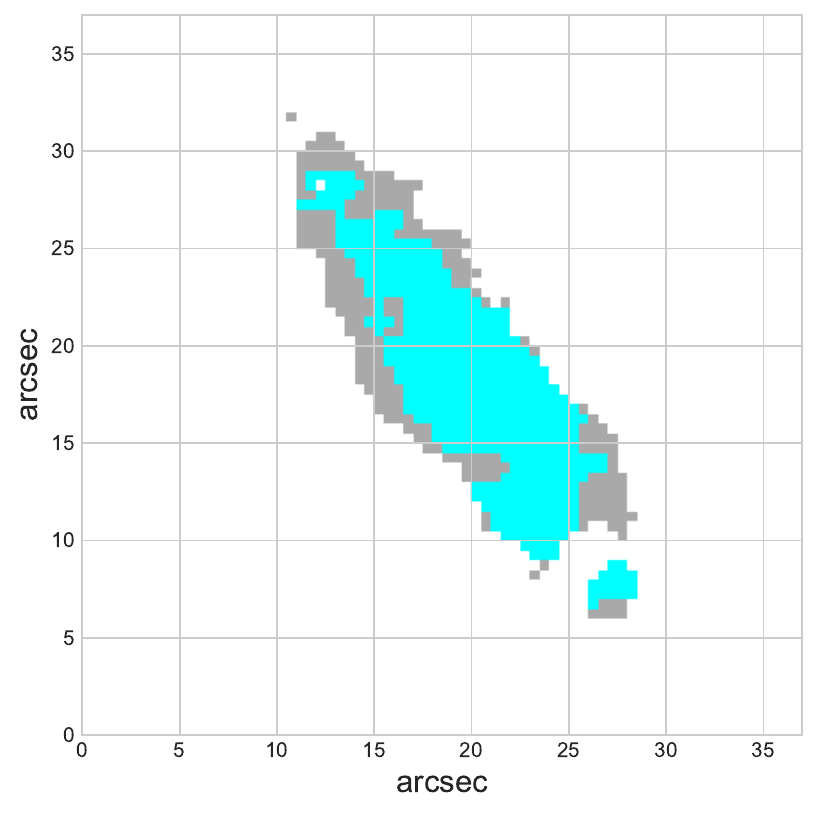} &
\includegraphics[width = 0.31\linewidth]{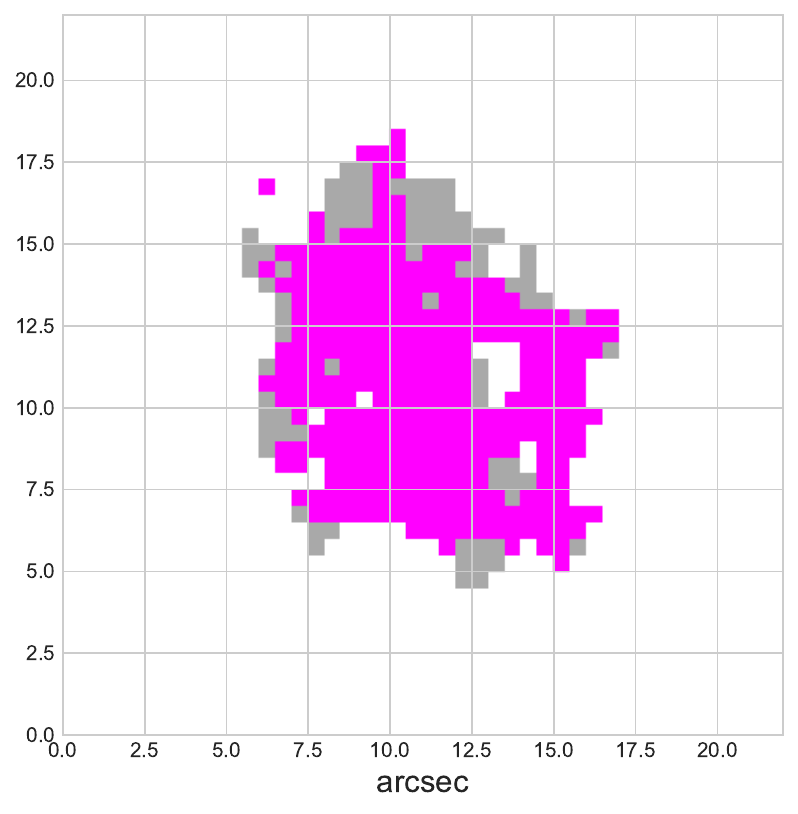} &
\includegraphics[width = 0.31\linewidth]{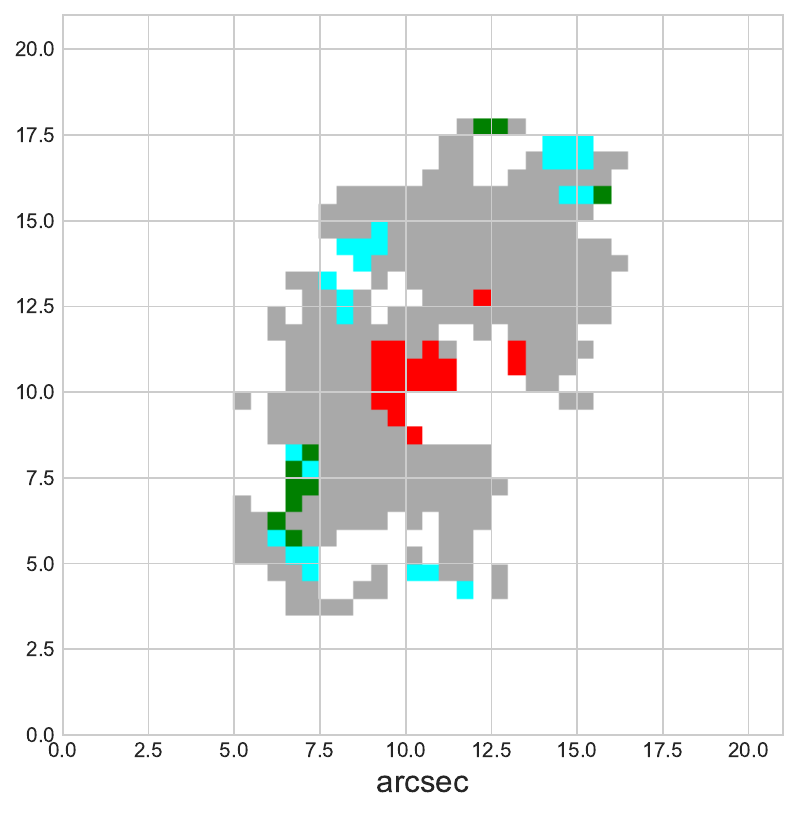}
\end{tabular}
\caption{Spatially resolved Baldwin–Phillips–Terlevich (BPT) classification maps for three representative CRDs (MaNGA IDs 1-54623, 1-584701, and 1-386322). Colors indicate the dominant ionization mechanism in each spaxel: star formation (cyan), composite (green), Seyfert (red), LINER (magenta), and ambiguous (gray).}
\label{fig: classifications}
\end{figure*}

MaNGA's observed wavelength range (3600\;-\;10,300\AA) covers all the strong nebular lines used in the standard BPT (Baldwin, Phillips \& Terlevich, \citet{BPT1981}) diagram. Specific to MaNGA, Marvin's \citep{Marvin2019} BPT function also has the ability to map the classification masks to the corresponding spaxels in the galaxy to further visualize which regions correspond to what mechanism. For the following investigation, the hydrogen lines S/N limit was assigned to 5, the OIII, SII, and NII S/N limit was assigned to 3, and the OI S/N limit was assigned to 1 (referred to as the ``\{5,3,1\}" configuration henceforth). Because of its abundance, we expect the hydrogen emission lines to be very bright, therefore we can set this required S/N to be fairly large. Typically the standard S/N is 3 (the Marvin default), therefore this is adopted for OIII, SII, and NII. Finally, the  OI emission lines are typically dim, therefore we can set this to be fairly small. 

All ionization source classifications were made through visual inspection, being categorized as either star-formation (SF), active galactic nuclei (AGN, both LINERs and Seyferts), ambiguous, or none. If multiple sources were present in the galaxy with no clear dominant source or if there was a significant amount of ambiguous sources, that galaxy was classified as ambiguous. Three galaxies classified as SF, AGN, and ambiguous can be seen in Fig.~\ref{fig: classifications}, for reference.

\begin{figure*} [t]
\centering
\begin{tabular}{ccc}
\includegraphics[width = 0.34\linewidth]{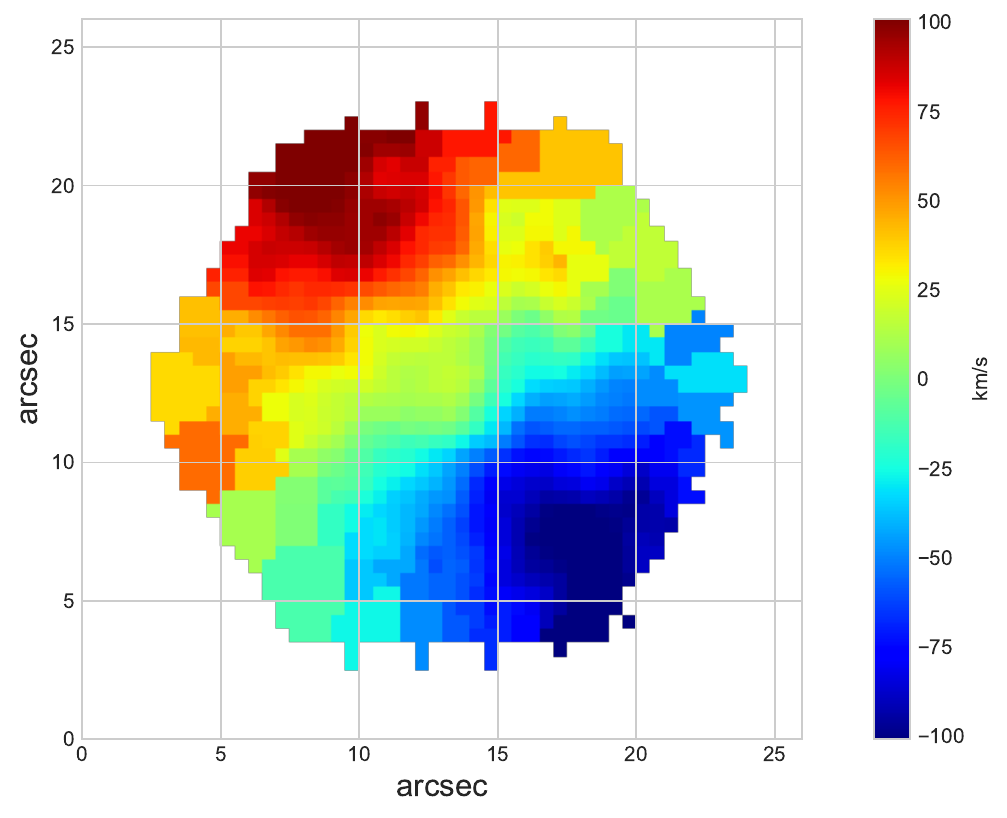} &
\includegraphics[width = 0.325\linewidth]{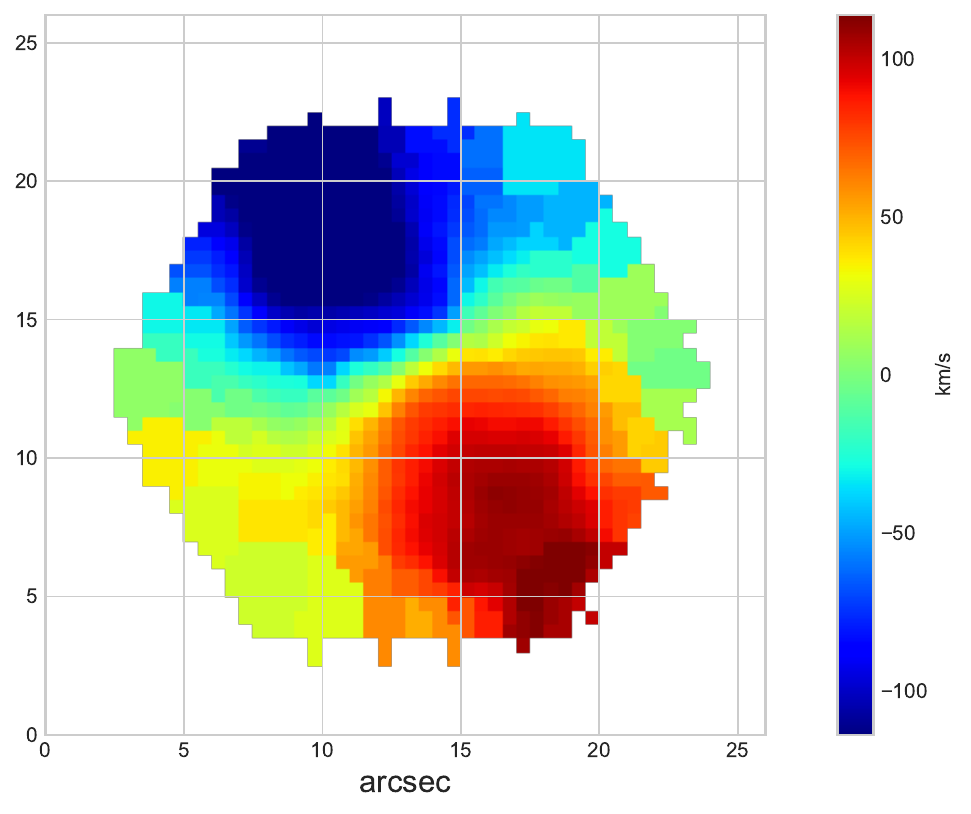} &
\includegraphics[width = 0.27\linewidth]{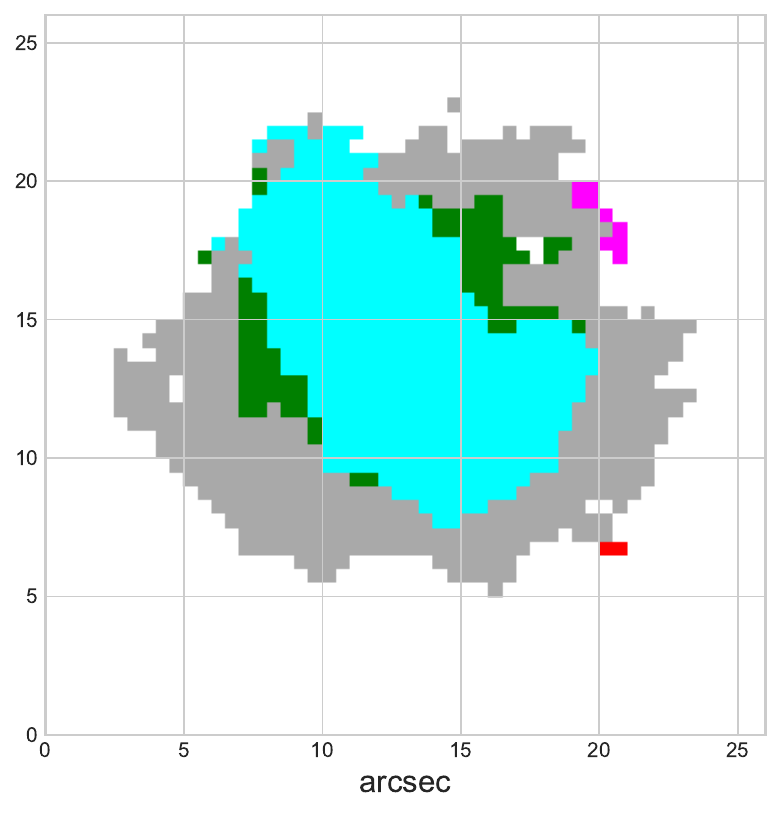}
\end{tabular}
\caption{Stellar velocity ($V_\star$; left), H$\alpha$ velocity (center), and spatially resolved Baldwin–Phillips–Terlevich (BPT) classification map (right) for MaNGA ID 1-600853. In the BPT panel, cyan spaxels indicate star-forming (SF) regions, green indicate composite emission, and pink indicate AGN/LINER-like excitation.}
\label{fig: less-dom galaxy}
\end{figure*}

Starting from the 126 extended CRD sample found in Section~\ref{sec: results}, visual inspection of the mapped BPT diagrams revealed 49 galaxies are dominated by SF ionization sources, 22 galaxies are dominated by AGN ionization sources, 10 galaxies are dominated by ambiguous ionization sources, and 45 galaxies have no detected emission lines with our set \{5,3,1\} scheme. Of the 49 CRDs dominated by SF, 29 CRDs exhibit a gaseous disk clearly corotating with the inner stellar disk, and 20 CRDs exhibit a gaseous disk clearly corotating with the outer stellar disk. Further, of these 20 CRDs with the exterior-corotating gaseous disk, 14 only had a single stellar disk visible in $V_\star$, suggesting this gaseous disk is corotating with the more-dominant stellar disk. Conversely, all 29 CRDs with the gaseous disk corotating inner stellar disk exhibited two visible stellar disks, with the inner stellar disk being less dominant than the outer stellar disk (i.e. the amplitude of the inner disk's LOSVD was less than the outer disk's). An example galaxy can be seen in Fig.~\ref{fig: less-dom galaxy}, where from $V_\star$ it is clear the outer stellar disk is much more prominent than the inner stellar disk, however the H$\alpha$ map suggests the gaseous disk is counter-rotating the more prominent stellar disk, ergo co-rotating with the inner stellar disk. 

The large fraction of galaxies without detected emission lines is primarily a consequence of the emission-line S/N requirement of the MaNGA DAP, rather than the absence of gas entirely. Our BPT analysis therefore only applies to galaxies with sufficiently strong nebular emission. As a result, the data presented here test whether the dominant ionization mechanism differs among \textit{emission-line} CRDs and non-CRDs; they do not constrain whether CRDs are more or less likely to host detectable ionized gas in the first place.

\subsection{Matching Set of non-CRDs}

To investigate the star-forming proportion of CRDs as compared to non-CRDs, we created a matching set of non-CRD galaxies and classified their dominant photoionization sources. This matching sample acts as a control group. By selecting galaxies with similar global properties, any differences in ionization source should be attributed primarily to the presence or absence of a CRD.

To create this matching set, we required the matching galaxies to have a similar mass, be at a similar redshift, and have a similar effective radius (being a proxy for size). Matching in stellar mass is particularly important because the fraction of star-forming and AGN galaxies is known to depend strongly on galaxy mass \citep{Kauffmann2003, Peng2010, Best2005}. For each CRD, a galaxy was considered to be a match if its own mass and effective radius were within 20\% of the CRDs mass and effective radius. For the redshift estimate, the total range of redshifts of all available galaxies in DR17 was broken into sixths, and the matching galaxy was required to be within this sixth-spacing-parameter of the redshift of the CRD. These percentiles were valid for 110 CRDs; for 13 of the remaining 15 CRDs the mass and effective radius percentiles were extended to 35\% and the redshift-spacing-parameter was brought down to a fourth, and for MaNGA IDs 1-373136 and 1-594318, being abnormally low and high mass galaxies respectively, the mass percentile was extended to 50\% (with only one matching galaxy being found for both CRDs). If a CRD was found to have multiple matching non-CRDs, one was selected at random. 

With this matching set, using the \{5,3,1\} configuration, visual inspection of the mapped BPT diagrams revealed 51 galaxies are dominated by SF ionization sources, 18 galaxies are dominated by AGN ionization sources, 14 galaxies are dominated by ambiguous ionization sources, and 43 galaxies with no emission line data. These statistics and the CRD statistics can be seen in Table~\ref{tab: BPT stats}. 

	\begin{table}[h] 
	\centering 
	\begin{tabular}{c||c|c|c|c} \hline
		& \textbf{\ \ SF \ \ } & \textbf{ \ AGN \ \quad} & \textbf{\ Ambig. \ } & \textbf{ \ None \ } \\ \hline \hline
		\textbf{\ \ CRD\ \ } & 49 & 22 & 10 & 45 \\
		\textbf{\ \ non-CRD\ \ } & 51 & 18 & 14 & 43 \\ \hline
	\end{tabular}
	\caption{Dominant BPT ionization classifications for the 126 confirmed CRDs and their matched non-CRD control sample using the {5,3,1} S/N configuration.}
	\label{tab: BPT stats}
	\end{table}

The distributions are remarkably similar, with only a two galaxy discrepancy between SF-dominated galaxies and a four galaxy discrepancy between AGN- and ambiguous-dominated galaxies. For the sample size, these differences are not statistically significant. Physically, this null result is plausible. Hydrogen recombination lines such as H$\alpha$ trace massive O-type stars and therefore probe star formation on timescales of $\lesssim$10 Myr \citep{Kennicutt2012}. If the counter-rotating stellar disk formed several hundred Myr to Gyr ago, the associated star-formation signatures may have already faded. Consequently, the absence of a measurable difference in dominant ionization sources does not contradict gas-accretion formation scenarios; instead, it indicates that BPT diagnostics are insensitive to the long timescales over which CRDs persist. This result therefore suggests that the dominant photoionization source alone is not a reliable proxy for the presence of counter-rotation.

The observed mixture of gas corotation with either the inner or outer stellar disk indicates that ionized gas in CRDs does not consistently align with a single stellar component. This behavior is consistent with external gas accretion scenarios, in which newly accreted gas can settle into either disk configuration before forming stars.

\section{Conclusion} \label{sec: conclusion}

In this paper, we introduce \FindingCRDs, a tool designed to identify counter-rotating stellar disks in IFU spectroscopy. The algorithm parameters were trained on the 64 CRDs identified by \cite{Bevacqua_2021} and empirically tuned to successfully recover the training sample. Applying \FindingCRDs to MaNGA DR17 flagged 1801 candidate galaxies, reducing the number of galaxies requiring visual inspection by 85\%. Through subsequent visual inspection, we identified 126 galaxies with clear counter-rotating stellar disks and an additional 143 galaxies (``mCRDs") whose counter-rotation signatures are more ambiguous. Because visual classification is inherently subjective, we provide both samples to enable flexible sample selection for future studies and to establish the largest currently available CRD sample.

Using the confirmed CRD sample, we investigated dominant photoionization sources via BPT diagnostics and compared the results to a matched control sample of early-type galaxies. The control galaxies were matched within 20\% in stellar mass, effective radius, and redshift (with only 15 CRDs requiring slightly relaxed thresholds). We find no statistically significant differences in emission-line classifications between CRDs and the control sample. These results imply that:

\begin{enumerate}
\item BPT emission-line diagnostics alone cannot reliably identify counter-rotating stellar disks.
\item The observed mixture of gas corotation with either stellar disk is consistent with external gas accretion scenarios in which newly acquired gas can settle into either rotational configuration.
\item Counter-rotation does not correspond to a distinct present-day photoionization signature.
\end{enumerate}

By enabling efficient identification of counter-rotating stellar disks in large IFU surveys, \FindingCRDs substantially expands the available sample and provides a foundation for future statistical studies of their formation and dynamical evolution.

\begin{deluxetable*}{ccccccccc}
\caption{Physical and kinematic properties of the 126 confirmed CRDs identified with \FindingCRDs. Columns list: (1) MaNGA ID; (2) TTYPE morphological parameter from the MDLM-VAC; (3) stellar mass; (4) effective radius; (5) logarithmic velocity dispersion within one effective radius; (6) kinematic detection criterion (CR, 2$\sigma$, or both); (7) stellar kinematic position angle; (8) H$\alpha$ kinematic position angle; and (9) qualitative alignment of the gaseous disk with respect to the inner or outer stellar disk.}
\label{tab: dr17 info}
\tablewidth{0pt}
\tabletypesize{\scriptsize}
\tablehead{
\colhead{MaNGA ID} &
\colhead{TTYPE} &
\colhead{$\log_{10}(M_\star/M_\odot)$} &
\colhead{$R_{\rm e}$ [kpc]} &
\colhead{$\log_{10}\sigma_{\rm e}$ [km s$^{-1}$]} &
\colhead{Detection} &
\colhead{PA$_\star$ [$^\circ$]} &
\colhead{PA$_{\rm gas}$ [$^\circ$]} &
\colhead{Gas Alignment} \\
\colhead{(1)} & \colhead{(2)} & \colhead{(3)} & \colhead{(4)} &
\colhead{(5)} & \colhead{(6)} & \colhead{(7)} &
\colhead{(8)} & \colhead{(9)}
}
\startdata
1-115097 & -0.9 & 9.99 & 3.29 & 2.05 & 2$\sigma$ & 57.5 & 76.0 & Inner \\
1-36923$^\S$ & 5.6 & 9.21 & 6.41 & 1.73 & CR & 152.0 & 96.0 & Outer \\
1-38347 & -1.3 & 10.55 & 3.05 & 2.34 & CR & 116.0 & 150.0 & Outer \\
1-40719 & -1.0 & 10.48 & 5.26 & 2.29 & CR & 87.0 & 98.0 & Inner \\
1-339061 & -1.6 & 10.14 & 6.32 & 2.18 & 2$\sigma$ & 138.0 & 162.0 & Inner \\
1-44047 & -1.0 & 9.92 & 1.46 & 2.11 & CR + 2$\sigma$ & 59.0 & 21.0 & Outer \\
1-44483 & 2.8 & 9.95 & 4.00 & 2.11 & CR + 2$\sigma$ & 97.0 & 93.0 & Outer \\
1-556514 & -2.1 & 10.68 & 6.02 & 2.3 & CR & 11.0 & - & ? \\
1-584701 & -0.0 & 10.54 & 8.05 & 2.15 & 2$\sigma$ & 117.0 & 65.0 & Misaligned \\
1-232087$^\S$ & 1.7 & 10.04 & 3.47 & 2.11 & CR + 2$\sigma$ & 65.0 & 62.0 & Outer \\
1-38543 & -0.2 & 10.10 & 2.23 & 2.13 & 2$\sigma$ & 93.0 & 93.5 & Inner \\
1-41971 & -1.7 & 10.90 & 9.74 & 2.34 & CR & 82.0 & - & ? \\
1-255220 & -1.1 & 9.57 & 1.12 & 1.94 & 2$\sigma$ & 158.5 & 114.0 & Inner \\
1-199775 & -0.8 & 9.96 & 3.08 & 1.89 & 2$\sigma$ & 165.5 & 168.5 & Inner \\
1-266244$^\S$ & 3.3 & 8.97 & 1.46 & 1.7 & 2$\sigma$ & 51.5 & 49.0 & Outer \\
1-251783 & -2.4 & 10.04 & 2.44 & 1.98 & CR & 178.5 & - & ? \\
1-251788 & -1.5 & 9.96 & 3.13 & 2.05 & CR + 2$\sigma$ & 178.5 & 36.0 & Outer \\
1-419236$^\S$ & 4.4 & 9.05 & 1.51 & 1.82 & CR & 152.0 & 78.5 & Inner \\
1-419257 & 2.1 & 9.99 & 2.86 & 2.06 & 2$\sigma$ & 79.0 & 75.5 & Inner \\
1-147496$^\S$ & 5.4 & 9.26 & 3.01 & 1.73 & CR & 35.5 & 99.0 & Outer \\
1-274545$^\S$ & 2.5 & 9.43 & 2.14 & 1.86 & CR & 106.5 & 133.0 & Inner \\
1-167044 & -2.7 & 10.86 & 8.21 & 2.39 & CR & 55.0 & - & ? \\
1-166613 & -0.7 & 10.88 & 7.72 & 2.34 & CR & 65.0 & 58.0 & Inner \\
1-210728 & -0.3 & 10.17 & 3.94 & 2.1 & CR & 19.5 & - & ? \\
1-248869 & -1.3 & 10.64 & 4.84 & 2.32 & 2$\sigma$ & 116.5 & 84.0 & Inner \\
1-136248 & -0.4 & 10.36 & 6.45 & 2.13 & CR + 2$\sigma$ & 73.5 & 84.5 & Outer \\
1-634718$^\S$ & 0.1 & 9.23 & 1.38 & 1.7 & CR + 2$\sigma$ & 35.5 & 44.0 & Inner \\
1-564483 & -2.3 & 10.62 & 5.81 & 2.3 & CR & 27.0 & 30.5 & Outer \\
1-179561 & -1.3 & 9.80 & 1.88 & 1.97 & CR + 2$\sigma$ & 101.0 & 102.0 & Inner \\
1-180900 & -0.5 & 10.49 & 3.34 & 2.14 & CR & 88.0 & - & ? \\
1-26197 & -0.8 & 10.79 & 9.10 & 2.29 & CR & 34.5 & 41.0 & Inner \\
1-378903$^\S$ & 7.1 & 8.79 & 6.66 & 1.77 & CR & 133.0 & 117.0 & Outer \\
1-163594 & -1.7 & 10.85 & 4.32 & 2.38 & CR & 51.0 & 159.5 & Misaligned \\
1-279073 & -0.2 & 10.56 & 5.76 & 2.34 & CR & 13.5 & 152.0 & Outer \\
1-235983$^\S$ & -1.8 & 9.62 & 1.01 & 2.04 & CR & 21.5 & 146.5 & Misaligned \\
1-457547 & 1.3 & 9.66 & 1.91 & 2.01 & 2$\sigma$ & 155.5 & 152.0 & Inner \\
1-174947 & 2.8 & 10.45 & 7.66 & 2.26 & CR + 2$\sigma$ & 66.0 & 65.0 & Outer \\
1-278079 & -2.1 & 11.26 & 12.65 & 2.47 & CR + 2$\sigma$ & 3.5 & - & ? \\
1-188530 & -1.1 & 10.37 & 3.32 & 2.31 & CR & 167.5 & 27.0 & Inner \\
1-149172$^\S$ & 2.0 & 8.88 & 1.05 & 1.7 & CR & 156.0 & 174.0 & Inner \\
1-148987 & 2.0 & 10.05 & 3.04 & 2.12 & 2$\sigma$ & 87.5 & 93.0 & Outer \\
1-94773$^\S$ & 0.4 & 9.90 & 1.52 & 1.98 & 2$\sigma$ & - & 58.5 & ? \\
1-94690 & 2.4 & 9.98 & 3.15 & 2.06 & 2$\sigma$ & 52.0 & 61.5 & Inner \\
1-323766 & -2.2 & 9.67 & 1.10 & 1.92 & CR & 129.0 & 119.5 & Inner \\
1-323764$^\S$ & 5.0 & 9.43 & 2.17 & 1.79 & CR & 93.0 & 36.0 & Outer \\
1-269227 & -2.7 & 10.21 & 2.34 & 2.34 & 2$\sigma$ & 46.0 & - & ? \\
1-560758$^\S$ & 0.9 & 9.60 & 1.88 & 1.71 & CR & 130.0 & 130.5 & Misaligned \\
1-549076 & -1.7 & 10.78 & 6.85 & 2.43 & 2$\sigma$ & 68.0 & - & ? \\
1-109275 & 0.3 & 9.80 & 2.56 & 1.93 & CR & 155.5 & 179.0 & Inner \\
1-382889 & 2.7 & 10.08 & 3.27 & 2.12 & 2$\sigma$ & 67.5 & 75.0 & Inner \\
1-387045$^\S$ & -0.1 & 9.65 & 1.31 & 1.95 & CR & - & 22.0 & ? \\
1-593328 & -1.5 & 11.10 & 13.20 & 2.41 & CR + 2$\sigma$ & 31.5 & 36.0 & Inner \\
1-322291$^\S$ & 0.3 & 9.89 & 4.51 & 1.92 & CR & 13.0 & - & ? \\
1-633000 & 0.4 & 9.76 & 1.75 & 1.98 & CR + 2$\sigma$ & 26.0 & 81.0 & Inner \\
1-322172$^\S$ & 4.5 & 8.84 & 1.78 & 1.63 & CR & 178.5 & 147.5 & Misaligned \\
1-623416 & -1.2 & 9.89 & 1.53 & 2.15 & CR & 107.5 & - & ? \\
\enddata
\end{deluxetable*}

\addtocounter{table}{-1} 
\begin{deluxetable*}{ccccccccc}
\tablecaption{(continued)}
\label{tab: dr17 info}
\tablewidth{0pt}
\tabletypesize{\scriptsize}
\tablehead{
\colhead{MaNGA ID} &
\colhead{TTYPE} &
\colhead{$\log_{10}(M_\star/M_\odot)$} &
\colhead{$R_{\rm e}$ [kpc]} &
\colhead{$\log_{10}\sigma_{\rm e}$ [km s$^{-1}$]} &
\colhead{Detection} &
\colhead{PA$_\star$ [$^\circ$]} &
\colhead{PA$_{\rm gas}$ [$^\circ$]} &
\colhead{Gas Alignment} \\
\colhead{(1)} & \colhead{(2)} & \colhead{(3)} & \colhead{(4)} &
\colhead{(5)} & \colhead{(6)} & \colhead{(7)} &
\colhead{(8)} & \colhead{(9)}
}
\startdata
1-456884 & 0.5 & 9.24 & 1.07 & 1.86 & CR + 2$\sigma$ & 22.0 & - & ? \\
1-623710$^\S$ & -0.0 & 9.00 & 2.66 & 1.57 & CR & 44.0 & - & ? \\
1-537081 & -1.2 & 10.85 & 5.90 & 2.2 & 2$\sigma$ & 132.0 & 64.0 & Outer \\
1-280716 & -0.5 & 10.21 & 3.66 & 2.16 & CR & 148.0 & - & ? \\
1-77006$^\S$ & -2.0 & 9.84 & 1.39 & 2.02 & CR & 107.0 & 110.5 & Inner \\
1-182730 & -1.9 & 10.82 & 6.58 & 2.28 & CR & 73.5 & - & ? \\
1-181720 & -2.0 & 10.70 & 8.46 & 2.28 & CR & 63.5 & 101.0 & Inner \\
1-591054 & -0.6 & 11.06 & 21.42 & 2.36 & CR & 97.5 & - & ? \\
1-320614 & -2.9 & 11.17 & 12.66 & 2.44 & CR & 127.5 & - & ? \\
1-245235$^\S$ & 2.0 & 9.84 & 2.35 & 2.05 & CR + 2$\sigma$ & 100.5 & 104.5 & Inner \\
1-245176$^\S$ & 0.0 & 9.92 & 1.94 & 2.08 & 2$\sigma$ & 131.0 & 149.5 & Outer \\
1-559761 & -2.2 & 10.41 & 4.08 & 2.26 & 2$\sigma$ & 14.5 & - & ? \\
1-54623 & 4.2 & 9.64 & 4.54 & 1.83 & 2$\sigma$ & 50.0 & 39.0 & Outer \\
1-83451 & -2.6 & 10.92 & 12.68 & 2.36 & CR & 15.5 & - & ? \\
1-13736 & -1.3 & 9.77 & 1.87 & 1.91 & 2$\sigma$ & 75.0 & 75.5 & Inner \\
1-586725 & -1.2 & 10.09 & 2.25 & 2.15 & 2$\sigma$ & 173.0 & 172.5 & Inner \\
1-211925 & -0.8 & 11.30 & 18.51 & 2.49 & CR & 89.5 & 121.0 & Outer \\
1-566335 & 0.1 & 10.17 & 5.18 & 2.21 & CR + 2$\sigma$ & 96.0 & 96.0 & Outer \\
1-333770 & 0.4 & 10.87 & 9.14 & 2.25 & CR & 44.0 & - & ? \\
1-118005 & 2.1 & 9.97 & 4.80 & 1.87 & 2$\sigma$ & 111.0 & 119.5 & Outer \\
1-40771$^\S$ & -2.1 & 11.10 & 11.83 & 2.33 & CR & 4.0 & 28.0 & Outer \\
1-386932$^\S$ & 1.7 & 9.55 & 1.59 & 1.94 & 2$\sigma$ & 13.5 & 27.5 & Inner \\
1-300092 & 0.7 & 10.55 & 7.01 & 2.19 & CR & 154.0 & 158.5 & Outer \\
1-139814$^\S$ & 1.0 & 9.94 & 2.31 & 1.84 & 2$\sigma$ & 2.5 & 178.5 & Outer \\
1-245 & -0.2 & 10.73 & 6.03 & 2.35 & CR & 114.0 & - & ? \\
1-2880$^\S$ & 3.7 & 8.97 & 1.99 & 1.76 & 2$\sigma$ & 30.5 & 41.5 & Outer \\
12-193534$^\S$ & 4.5 & 8.84 & 1.78 & 1.72 & CR + 2$\sigma$ & 119.5 & 148.0 & Inner \\
1-113520 & -1.3 & 9.49 & 1.16 & 1.78 & CR + 2$\sigma$ & 117.5 & 6.5 & Misaligned \\
1-38062$^\S$ & -0.3 & 10.02 & 4.14 & 1.92 & CR + 2$\sigma$ & 145.5 & 126.0 & Outer \\
1-41258 & 0.1 & 10.57 & 7.75 & 2.25 & CR & 41.5 & 37.0 & Outer \\
1-37494 & -1.2 & 9.96 & 2.90 & 2.1 & 2$\sigma$ & 150.5 & 132.0 & Outer \\
1-137890 & 1.6 & 9.62 & 1.74 & 2.04 & CR + 2$\sigma$ & 76.0 & 61.0 & Inner \\
1-247630 & -0.4 & 10.51 & 6.13 & 2.16 & CR + 2$\sigma$ & 155.5 & 168.0 & Outer \\
1-635590 & -0.5 & 11.06 & 9.30 & 2.47 & 2$\sigma$ & 166.0 & - & ? \\
1-248410 & 0.4 & 9.28 & 1.45 & 1.85 & 2$\sigma$ & 15.5 & 168.5 & Outer \\
1-373136$^\S$ & 3.4 & 9.38 & 4.26 & 1.76 & CR & 143.5 & 86.5 & Inner \\
1-121871 & 0.9 & 9.92 & 2.82 & 1.91 & 2$\sigma$ & 85.5 & 137.5 & Outer \\
1-42247 & 0.3 & 9.40 & 1.46 & 1.87 & CR & 134.0 & 145.5 & Inner \\
1-297863 & -1.9 & 9.99 & 1.64 & 2.04 & 2$\sigma$ & 29.5 & - & ? \\
1-78381 & 1.0 & 10.86 & 12.55 & 2.23 & CR & 79.5 & 75.0 & Outer \\
1-135767 & -2.6 & 10.42 & 11.60 & 2.13 & CR & 103.0 & 103.5 & Outer \\
1-275185 & -2.0 & 10.81 & 6.49 & 2.39 & CR & 63.5 & 145.5 & Misaligned \\
1-29809 & 0.1 & 9.31 & 1.39 & 1.69 & 2$\sigma$ & 155.0 & 133.5 & Inner \\
1-600853 & 1.4 & 10.18 & 4.78 & 2.05 & 2$\sigma$ & 34.5 & 28.5 & Inner \\
1-386154 & 1.6 & 9.67 & 1.92 & 1.87 & 2$\sigma$ & 153.5 & 148.0 & Inner \\
1-42082 & -2.0 & 10.57 & 3.86 & 2.32 & CR & 96.0 & 83.5 & Inner \\
1-401708 & 3.6 & 10.70 & 10.30 & 2.29 & 2$\sigma$ & 121.5 & 120.5 & Outer \\
1-421145 & -0.9 & 9.88 & 2.27 & 1.98 & 2$\sigma$ & 178.5 & - & ? \\
1-418090 & -1.3 & 10.70 & 6.36 & 2.2 & CR & 149.0 & - & ? \\
1-25632 & 1.7 & 10.25 & 5.17 & 1.99 & 2$\sigma$ & 167.0 & 168.5 & Outer \\
1-352130$^\S$ & 4.1 & 8.73 & 2.67 & 1.51 & CR & 47.5 & 84.5 & Outer \\
1-323886$^\S$ & 2.3 & 9.78 & 2.70 & 2.0 & 2$\sigma$ & 62.0 & 76.0 & Outer \\
1-594318 & -2.1 & 11.44 & 16.42 & 2.53 & 2$\sigma$ & 125.5 & 115.5 & Inner \\
1-386322 & 1.3 & 9.67 & 1.85 & 1.95 & 2$\sigma$ & 154.0 & 153.5 & Outer \\
1-189376$^\S$ & 1.6 & 9.65 & 2.03 & 1.78 & 2$\sigma$ & 14.5 & 17.0 & Inner \\
1-389466$^\S$ & 0.1 & 9.68 & 1.52 & 1.97 & CR & 113.0 & 129.5 & Inner \\
1-456731 & -0.2 & 9.27 & 0.94 & 1.85 & CR & 99.0 & 159.5 & Inner \\
1-522189 & -2.0 & 9.94 & 1.87 & 1.81 & CR & 80.5 & 119.0 & Misaligned \\
1-301561$^\S$ & 5.3 & 9.24 & 3.45 & 1.8 & 2$\sigma$ & 106.0 & 99.0 & Outer \\
1-546250 & -1.6 & 10.85 & 9.03 & 2.34 & CR & 19.5 & 96.5 & Outer \\
\enddata
\end{deluxetable*}

\addtocounter{table}{-1} 
\begin{deluxetable*}{ccccccccc}[t]
\tablecaption{(continued)}
\label{tab: dr17 info}
\tablewidth{0pt}
\tabletypesize{\scriptsize}
\tablehead{
\colhead{MaNGA ID} &
\colhead{TTYPE} &
\colhead{$\log_{10}(M_\star/M_\odot)$} &
\colhead{$R_{\rm e}$ [kpc]} &
\colhead{$\log_{10}\sigma_{\rm e}$ [km s$^{-1}$]} &
\colhead{Detection} &
\colhead{PA$_\star$ [$^\circ$]} &
\colhead{PA$_{\rm gas}$ [$^\circ$]} &
\colhead{Gas Alignment} \\
\colhead{(1)} & \colhead{(2)} & \colhead{(3)} & \colhead{(4)} &
\colhead{(5)} & \colhead{(6)} & \colhead{(7)} &
\colhead{(8)} & \colhead{(9)}
}
\startdata
1-116100$^\S$ & 4.1 & 9.55 & 4.28 & 1.78 & CR & 37.5 & 52.5 & Outer \\
1-519907 & 0.6 & 9.76 & 1.47 & 2.05 & 2$\sigma$ & 108.5 & 99.5 & Misaligned \\
1-244115 & 0.0 & 9.87 & 2.18 & 2.13 & 2$\sigma$ & 113.0 & 103.5 & Inner \\
1-282035$^\S$ & 2.7 & 9.12 & 1.20 & 1.73 & CR & 88.5 & 20.5 & Misaligned \\
1-418023$^\S$ & -1.5 & 9.49 & 0.93 & 1.81 & 2$\sigma$ & 45.5 & 99.0 & Outer \\
1-44722 & -1.7 & 9.82 & 6.05 & 2.05 & CR & 80.5 & - & ? \\
12-84617$^\S$ & 4.5 & 8.83 & 1.54 & 1.67 & CR & 47.0 & 58.5 & Outer \\
1-113698$^\S$ & 2.2 & 8.72 & 1.11 & 1.73 & CR & 33.0 & 50.0 & Inner \\
1-298940 & -1.7 & 10.08 & 3.68 & 2.06 & CR & 45.5 & - & ? \\
1-76425 & 3.2 & 10.47 & 6.84 & 2.16 & 2$\sigma$ & 138.5 & 119.5 & Inner \\ \\
\enddata
\end{deluxetable*}

\begin{deluxetable*}{cc|cc|cc|cc} [b]
\caption{MaNGA IDs and primary kinematic detection feature for the 143 “mCRD” (borderline) galaxies. Detection indicates whether the system was flagged via counter-rotation (CR), a double-peaked velocity dispersion profile (2$\sigma$), or both.}
\label{tab: mCRD}
\tablewidth{0pt}
\tabletypesize{\scriptsize}
\tablehead{
\colhead{MaNGA ID} &
\colhead{Detection} &
\colhead{MaNGA ID} &
\colhead{Detection} &
\colhead{MaNGA ID} &
\colhead{Detection} &
\colhead{MaNGA ID} &
\colhead{Detection} 
}
\startdata
1-177529 & $2\sigma$ & 1-179096 & $2\sigma$ & 1-179600 & CR & 1-179355 & CR \\
1-113242 & CR & 1-116340 & CR & 1-22414 & CR & 1-42070 & CR \\
1-36890 & CR & 1-38008 & $2\sigma$ & 1-109392 & CR & 1-41072 & CR \\
1-378276 & CR & 1-556514 & CR & 1-145741 & $2\sigma$ & 1-145871 & CR \\
1-52554 & CR & 1-137853 & CR & 1-495351 & CR & 1-285144 & CR \\
1-235270 & CR & 1-234980 & CR & 1-575741 & CR & 1-402656 & CR \\
1-491140 & $2\sigma$ & 1-216951 & $2\sigma$ & 1-254691 & CR & 1-275185 & CR \\
1-166739 & CR & 1-284552 & CR & 1-91550 & CR & 1-93551 & CR \\
1-114923 & CR & 1-23980 & CR & 1-29809 & $2\sigma$ & 1-600853 & $2\sigma$ \\
1-378910 & CR & 1-379490 & CR & 1-605161 & CR & 1-51949 & CR \\
1-218230 & CR + $2\sigma$ & 1-148949 & CR & 1-188177 & $2\sigma$ & 1-149507 & CR \\
1-373351 & CR & 1-272047 & CR & 1-594335 & CR & 1-632653 & $2\sigma$ \\
1-631890 & CR & 1-318513 & $2\sigma$ & 1-318069 & $2\sigma$ & 1-633775 & $2\sigma$ \\
1-153938 & CR & 1-178027 & $2\sigma$ & 1-42660 & $2\sigma$ & 1-153284 & CR \\
1-386154 & $2\sigma$ & 1-298867 & CR & 1-41918 & CR & 1-42082 & CR \\
1-456974 & CR & 1-246484 & $2\sigma$ & 1-456929 & $2\sigma$ & 1-456672 & CR \\
1-456299 & CR & 1-456714 & CR & 1-456616 & CR & 1-522651 & CR \\
1-376116 & CR & 1-153696 & $2\sigma$ & 1-136508 & CR & 1-153127 & $2\sigma$ \\
1-50622 & CR & 1-412352 & CR & 1-76813 & CR & 1-568566 & CR \\
1-456645 & $2\sigma$ & 1-320459 & CR & 1-1243 & CR & 1-325250 & CR \\
1-497670 & CR & 1-116700 & CR & 1-118491 & CR & 1-400885 & $2\sigma$ \\
1-10263 & $2\sigma$ & 1-66669 & $2\sigma$ & 1-401708 & $2\sigma$ & 1-273275 & $2\sigma$ \\
1-639484 & $2\sigma$ & 1-120019 & CR & 1-180392 & CR & 1-38612 & CR \\
1-463943 & CR & 1-481264 & CR & 1-558248 & CR & 1-277289 & $2\sigma$ \\
1-255963 & $2\sigma$ & 1-258311 & $2\sigma$ & 1-196580 & $2\sigma$ & 1-251269 & $2\sigma$ \\
1-261032 & $2\sigma$ & 1-147787 & $2\sigma$ & 1-255090 & CR & 1-274440 & $2\sigma$ \\
1-93908 & $2\sigma$ & 1-351511 & $2\sigma$ & 1-176829 & $2\sigma$ & 1-24132 & CR \\
1-72913 & CR & 1-217854 & CR & 1-298727 & CR + $2\sigma$ & 1-174074 & CR \\
1-270072 & CR & 1-265180 & CR & 1-42615 & CR & 1-594053 & CR \\
1-295411 & CR & 1-605544 & CR & 1-605510 & CR & 1-234172 & $2\sigma$ \\
1-96324 & CR & 1-270492 & CR & 1-42031 & CR & 12-193481 & $2\sigma$ \\
1-177635 & CR + $2\sigma$ & 1-178130 & $2\sigma$ & 1-252094 & $2\sigma$ & 1-278490 & $2\sigma$ \\
1-147514 & $2\sigma$ & 1-419380 & $2\sigma$ & 1-604826 & $2\sigma$ & 1-626395 & $2\sigma$ \\
1-195337 & $2\sigma$ & 1-543599 & $2\sigma$ & 1-135244 & $2\sigma$ & 1-418023 & $2\sigma$ \\
1-318146 & $2\sigma$ & 1-43022 & $2\sigma$ & 1-147488 & $2\sigma$ & 1-180981 & $2\sigma$ \\
1-265750 & $2\sigma$ & 1-246298 & $2\sigma$ & 1-195349 & $2\sigma$ \\ \\
\enddata
\end{deluxetable*}

\twocolumngrid
\bibliographystyle{aasjournal}
\bibliography{sample631}{}

\end{document}